\documentclass[12pt]{iopart}
\eqnobysec
\usepackage{graphicx}
\usepackage{color}

\newcommand{\be}{\begin{equation}}
\newcommand{\ee}{\end{equation}}

\newcommand{\x}{{\mathbf x}}

\newcommand{\J}{{\mathbf J}}

\newcommand{\B}{\mathbf{B}}

\newcommand{\V}{{\mathbf V}}

\newcommand{\M}{{\mathbf M}}

\newcommand{\Id}{{\mathbf I_d}}

\newcommand{\der}{\partial}

\begin{document}

\title{Energy transition density of driven chaotic systems: A compound trace formula}

\author {Alfredo M. Ozorio de Almeida\footnote{ozorio@cbpf.br}}
\address{Centro Brasileiro de Pesquisas Fisicas,
Rua Xavier Sigaud 150, 22290-180, Rio de Janeiro, R.J., Brazil}

\begin{abstract}

Oscillations in the probability density of quantum transitions of the eigenstates of a chaotic Hamiltonian within classically narrow energy ranges
have been shown to depend on closed compound orbits. These are formed by a pair of orbit segments, one in the energy shell of the original Hamiltonian
and the other in the energy shell of the driven Hamiltonian, with endpoints which coincide. Viewed in the time domain, 
the same pair of trajectory segments arises in the semiclassical evaluation of the trace of a compound propagator: 
the product of the complex exponentials of the original Hamiltonian and of its driven image. 
It is shown here that the probability density is the double
Fourier transform of this trace, so that the closed compound orbits emulate the role played by periodic orbits in Gutzwiller's trace formula 
in its semiclassical evaluation. The phase of the oscillations with the energies or evolution parameters agree with thoses previously obtained, 
whereas the amplitude of the contribution of each closed compound orbit is more compact and independent of any feature of the Weyl-Wigner representation in which the calculation was carried out.

\end{abstract}

\maketitle

\section{Introduction}

Initial quantum states, which undergo manipulations in a laboratory, are usually chosen to be simple,
such as coherent states, or the eigenstates of an integrable Hamiltonian. Even so, there is in principle
no impediment to the unitary evolution of excited eigenstates of chaotic systems. It is true that they may
be hard to single out in a dense energy spectrum with no selection rules, but this may be a further reason
to focus on transitions between energy windows, which contain many eigenstates, though being classically narrow. 
Theoretically, this choice is motivated by the lack of an apt description of individual chaotic eigenstates, 
such as the semiclassical (SC) portrayal of the eigenstates of integrable systems, through
their Wigner functions \cite{Wigner,Ber77}.  

Different features of the energy transition probability density, or the transition density for short,
were recently studied in a pair of papers \cite{transI,transII}, here labled {\bf I} and {\bf II}.  
Whereas {\bf I} concentrated on classical aspects and the occurrence of caustic singularities in the transition density, the nature of its SC oscillations away from caustics was the focus of {\bf II}.
Furthermore, this second paper generalized the special kind of transitions, which were allowed in {\bf I}. 
Thus, the initial eigenstates may evolve under the action of arbitrary Hamiltonians, 
as long as this can be assumed to be a fast process in comparison with internal motions.

The transition density was calculated  in {\bf II} as a stationary phase approximation of
an exact integral, which portrays the trace of a pair of operators, represented as  
a pair of spectral Wigner functions. Each of these is a superposition of Wigner functions
of the many eigenstates to be found semiclassically within a classically narrow energy window. 
In {\bf I} the spectral Wigner function was even expressed as a path integral,  
but, instead, well established SC approximations of the spectral Wigner were employed \cite{Ber89,Report}. 
The full SC scenario combining the results of both papers is presented in {\bf II}, so that the present
contribution deals only with an improvement of the amplitudes previously obtained. 

The energy transition density is in principle an experimental property, so that it must be independent of the representation chosen to perform its calculation. It is thus pleasing that the phase of the oscillations, in the SC derivation of {\bf II} for the transition density, erase any memory of the Weyl-Wigner representation, depending only on the action of {\it closed compound orbits} formed by a pair of orbit segments, 
one in an energy shell of the initial system and the other in an energy shell of the driven system.  
On the other hand, the amplitude of each compound orbit contribution resulted in a cumbersome product of the features 
of the pair of spectral Wigner functions
that are selected by the stationary phase condition. Presumably it is feasible to simplify this amplitude by clever manipulations, but an alternative path is here presented.

The object of the present paper is to provide a new derivation of the transition density, that starts from the
trace of a pair of evolution operators and has a simple SC expression in terms of closed compound trajectories \cite{OzBro16}. In this time-domain, both the phase and the amplitude are free from the special features of the Wigner-Weyl representation. The main task is then to invert the dependence on the pair of times, characterizing the trajectory segments that form the compound trajectory, to a dependence on the pair os energies
that emerge as the integrals over times, performed again within the stationary phase approximation.
\footnote{In most cases, the term trajectory will be employed in the time-domain and orbit in the energy-domain.} 
It so happens that the classical characterization
of the compound orbits in the energy-domain is more intuitively accessible than in the time-domain, 
so that delicate navigation between both pictures is needed. Thus, one obtains a more comprehensive view of the structure of closed compound
orbits than in the original derivation in {\bf II}, as well as the bonus of a simpler expression for the amplitude of their contribution.

The present derivation clarifies the parallelism of the structure of the energy transition density to that of the the smoothed density of eigenenergies. The latter may also be expressed as a Fourier transform of the trace of the evolution operator and the semiclassical approximation
of Gutzwiller \cite{Gutz,Gutzbook} and of Balian and Bloch \cite{BalBloch}
is based on the classical periodic orbits in each energy shell. Actually, the periodic orbits are already present 
in the trace of the propagator itself and it is only in the energy domain that smoothing is needed for convergence.
In the present case of a compound trace combining two propagations, the semiclassical evaluation of the double Fourier transform
relies instead on the compound orbits joining segments on both chosen energy shells. The classical bases of the static single energy density 
and the dynamical transition denstity are thus parallel, depending on closed structures with one dimension. But, of course,
compound orbits are much less familiar than periodic orbits and their zoology then needs to be developed here.

The alternative exact formula for the transition density is presented in the following section in the form of a double Fourier transform
of the trace of a compound evolution operator. Section 3 recalls the SC approximation of this trace, derived in \cite{OzBro16}
in terms of closed compound trajectories. Then the Fourier integral is performed within the stationary phase approximation, hence producing the SC approximation to the energy transition density in terms of closed compound orbits. Section 4 constructs  Poincar\'e  sections
from which one establishes the existence of sequences of families of closed compound orbits parametrized by a pair of energies 
and hence sequences of closed compound trajectories parametrized by pairs of times, as assumed in section 3.
Finally, section 5 compares the simpler SC expression for the energy transition density presented in section 3 
with its previous derivation in {\bf II}.

\section{Energy transitions driven by general unitary operators}

Let us consider the probability density for the transition of a quantum system with Hamiltonian $\hat H$ from an energy range centred on $E$ with a width $\epsilon$ to an energy range centred on $E'$ with the same width. It is driven by a one parameter family of unitary operators,
\be
\hat{U}(\tau) = \e^{-i\tau{\hat\Lambda}/\hbar}~,
\ee
where $\hat\Lambda$ is the driving Hamiltonian and $\tau$ is the driving time. It was expressed in {\bf II} as
\be
\fl P_{EE'}(\tau, \epsilon)  
=  \frac{1}{(2\pi\hbar)^{N}} \int {\rm d}^{2N}\x ~ W_E(\x, \epsilon)~W_{E'}(\x,\epsilon|\tau) ~.
\label{PEE1}
\ee
Within the integral, each spectral Wigner function $W_E(\x, \epsilon)$ is defined directly as a superposition of the Wigner functions $W_k(\x)$, for individual eigenstates $|k\rangle$ of $\hat H$, with energy $E_k$
\cite{Ber89,Ber89b,Report}:
\be
W_E(\x,\epsilon) \equiv (2\pi\hbar)^N \sum_k \delta_\epsilon(E-E_k)~ W_k(\x) ~.
\label{spectral1}
\ee
They represent the spectral density operator
\be
{\hat \rho}_E(\epsilon) \equiv \sum_k \delta_\epsilon(E-E_k)~ |k\rangle \langle k|
\ee
in the classically narrow energy range $\epsilon$ centred on an energy E, being that 
\be
\delta_\epsilon(E)\equiv \frac{1}{\pi} ~ \frac{\epsilon}{\epsilon^2 + E^2}
\label{widelta}
\ee 
integrates as a Dirac $\delta$-function.
In its turn, the unitary evolution of a pure state, that is, the driven pure state
\be
|l\rangle\langle l|(\tau) \equiv \hat{U}(\tau)|l\rangle\langle l|\hat{U}(\tau)^\dagger 
\ee
is represented by the driven Wigner functions $W_k(\x|\tau)$, which leads to the spectral Wigner function of the driven system
\be
W_{E'}(\x,\epsilon|\tau) \equiv (2\pi\hbar)^N \sum_k \delta_\epsilon(E'-E_k)~ W_k(\x|\tau) ~.
\label{spectralev}
\ee

The alternative construction of the spectral Wigner function goes back to the {\it Weyl propagator} $V(\x,t)$, 
which represents the intrinsic unitary evolution operator
\be
\hat{V}(t) = \e^{-it{\hat H}/ \hbar}~,
\label{intev}
\ee
that is,
\be
W_{E}(\x,\epsilon) = {\rm Re} \int_0^\infty \frac{{\rm d}t}{\pi\hbar}~\exp\left[ \frac{it}{\hbar}(E+i\epsilon)\right]~ V(\x,t) ~, 
\label{spectral2}
\ee  
or, more conveniently here,
\be
W_{E}(\x,\epsilon) = \int_{-\infty}^\infty \frac{{\rm d}t}{2\pi\hbar}~\exp\left[ \frac{iEt}{\hbar}- \frac{\epsilon|t|}{\hbar}\right]~ V(\x,t) ~. 
\label{spectral3}
\ee  
On the other hand, the driven Hamiltonian generated by $\hat{U}(\tau)$,
\be
{\hat H}(\tau) \equiv \hat{U}(\tau){\hat H}~\hat{U}(\tau)^\dagger ~,
\label{evHam}
\ee
generates the driven intrinsic evolution operator
\be
\hat{V}(t|\tau) \equiv \hat{U}(\tau) ~\e^{-it{\hat H}/ \hbar}~\hat{U}(\tau)^\dagger = \exp\left[-\frac{it}{\hbar}\hat H(\tau)\right]~,
\ee
represented by the corresponding Weyl propagator $V(\x,t|\tau)$. Thus, the driven spectral Wigner function is expressed as
\be
W_{E'}(\x,\epsilon|\tau) = \int_{-\infty}^\infty \frac{{\rm d}t'}{2\pi\hbar}~
\exp\left[ \frac{iE't'}{\hbar}-\frac{\epsilon|t'|}{\hbar}\right]~ V(\x,t'|\tau) ~, 
\label{spectral4}
\ee 
Introducing these expressions in (2.2) leads to the transition density in the form 
\begin{eqnarray}
\fl P_{EE'}(\tau, \epsilon)  =  &\frac{1}{(2\pi\hbar)^{N}} \int {\rm d}^{2N}\x \\   \nonumber
&\left\{\int_{-\infty}^\infty \frac{{\rm d}t}{2\pi\hbar}~\exp\left[ \frac{iEt}{\hbar}- \frac{\epsilon|t|}{\hbar}\right] V(\x,t) \right\}  
\left\{\int_{-\infty}^\infty \frac{{\rm d}t'}{2\pi\hbar}~\exp\left[ \frac{iE't'}{\hbar}-\frac{\epsilon|t'|}{\hbar}\right] V(\x,t'|\tau) \right\} ~,
\label{PEE2}
\end{eqnarray}
so that, switching the integrations,
\begin{eqnarray}
\fl P_{EE'}(\tau, \epsilon) = \int_{-\infty}^\infty \frac{{\rm d}t}{2\pi\hbar} \int_{-\infty}^\infty \frac{{\rm d}t'}{2\pi\hbar}
 ~ \exp\left[ -\frac{\epsilon}{\hbar}(|t|+|t'|)\right] ~ \exp\left[ i\frac{(Et + E't')}{\hbar} \right] 
\\   \nonumber
\int\frac{ {\rm d}^{2N}\x}{(2\pi\hbar)^{N}} ~ V(\x, t)~ V(\x, t'|\tau).
\label{PEE3}
\end{eqnarray}

At this stage, it is recalled the property of the Wigner-Weyl representation (see e.g.  \cite{Report}) that
\be
\int\frac{ {\rm d}^{2N}\x}{(2\pi\hbar)^{N}} )~ V(\x, t)~ V(\x, t'|\tau) = {\rm  tr} ~\hat{V}(t)~\hat{V}(t'|\tau) ~,
\ee
that is, the trace of the product of operators, which may be interpreted as the single {\it compound evolution operator} 
\be
\hat{\V}(t, t') \equiv \hat{V}(t)~\hat{V}(t'|\tau) ~,
\ee 
as defined in \cite{OzBro16}. Hence, the alternative exact expression for the energy transition density; that is the basis for the present
SC approximations, is the double Fourier transform
\be  
\fl P_{EE'}(\tau, \epsilon) = \int_{-\infty}^\infty \frac{{\rm d}t~{\rm d}t'}{(2\pi\hbar)^2}
 ~ \exp\left[ \frac{i}{\hbar}(Et + E't') \right]
 ~ \exp\left[ -\frac{\epsilon}{\hbar}(|t|+|t'|)\right] ~ {\rm tr} ~ \hat{\V}(t,t').
\label{PEE4}
\ee
Since the integral runs over positive and negative times, all combinations of forward and backward evolutions 
for both families of operators are included in the integral. The simultaneous reversal of both times produces 
the complex conjugate, so that the integral is real.

\section{Semiclassical approximations}

The original discussion of compound evolution operators in the context of evolving quantum correlations
considered products of an arbitrary number of evolutions in the Wigner-Weyl representation. It turns out
that the  SC trace of a compound operator derived in \cite{OzBro16} depends on the action of 
{\it compound trajectories}: they join trajectory segments from each of the evolutions into  piecewise smooth
closed curves in phase space. Here we shall be concerned with just a pair of curves, one evolved by
the classical Hamiltonian $H(\x)$ in the positive or negative time $t$ and the other evolved by the driven Hamiltonian $H(\x|\tau)$
in the time $t'$. Thus, one defines $\S(t, t')$, the classical action of the compound trajectory,
such that the SC contribution to the trace of the compound trajectory was shown to be
\begin{equation}
{\rm tr } ~ \hat{\V}(t, t') \approx \frac{2^N}{|\det[\Id - \M(t, t')]|^{1/2}}\>\>
\exp \left[ \frac{i}{\hbar}(\S(t, t')+ \hbar \sigma)\right] ~.
\label{Uweyl}
\end{equation}

The linear approximation of the transformation generated by $H(\x)$ near the $t$-trajectory being defined by the {\it stability matrix},
$\M(t)$ and the stability matrix for the linearized motion near the $t'$-trajectory being $\M'(t')$, then the full stability matrix for the compound trajectory is simply
\be
\M(t, t') \equiv \M(t')\M(t) ~, 
\ee
whereas $\Id$ is the identity matrix. One should note that the discontinuity of the derivative of the compound trajectory
at the initial and final points of both the trajectory segments implies that generally there is no zero eigenvalue of the
full stability matrix, unlike smooth periodic trajectories. On the other hand, the determinant in the amplitude of \eref{Uweyl}
does not depend on the choice of the starting point between both trajectory endpoints.   
 
There may be multiple branches of the action, meeting along caustics in the $(t,t')$ space, where the semiclassical amplitude diverges,
that is, $[\det\Id - \M(t,t')]=0$.
The phase $\sigma$ is determined by the convergence of neighbouring paths.
(For convenience of notation, all such focal indices shall henceforth be denoted by the generic symbol $\sigma$, without specifying 
their generic values, which will alter under the transformations to be carried out.)
There are no caustics if both times $t$ and $t'$ are short \cite{Report}, for which $\sigma=0$, but usually there is an increment at a caustic.

To evaluate the integral \eref{PEE4} by stationary phase, one needs the time derivatives 
of the action of the compound trajectory. Within the Wigner-Weyl representation, this can be decomposed into a pair of centre actions \cite{Report}
\be
\S(t, t') = S(\x, t) + S(\x, t'|\tau) ~,
\ee
where $\x$ is the centre of the vector joining the pair of endpoints of the pair of segments that form the compound trajectory. This is the the action of a compound trajectory defined by a pair of times, which corresponds to the action of a compound orbit 
defined in {\bf II} by a pair of energies.
Then
\be
\fl \frac{\der \S}{\der t}(t, t') = \frac{\der S}{\der t}(\x, t) = -t~E(t) ~~~~  {\rm and} ~~~~ 
\frac{\der \S}{\der t'}(t, t') = \frac{\der S}{\der t'}(\x, t'|\tau) = -t' ~E(t'|\tau) ~,
\ee
where $E(t)$ and $E(t'|\tau)$ are the respective energies of the $t$-trajectory and the $t'$- trajectories. 
\footnote{It should be noted that other choices of generating function, such as the ones depending on initial and final positions, result in the same partial derivatives of the action of the compound trajectory.}
Therefore the stationary phase condition for the integrand of \eref{Uweyl} is simply 
$E(t)=E$ and $E(t'|\tau) = E'$. Once both these conditions are satisfied, one retrieves the energy action
of the compound orbit in {\bf II},
\be
\S(E,E') = \S(t(E,E'),~ t'(E,E'))+ E ~ t(E,E') +  E' ~ t'(E,E') ~,
\ee
which is just the symplectic area within the compound orbit, illustrated in Fig. 1.

The assumption that one disposes of a two-parameter family of compound trajectories in time,
which are a function of the pair of energy parameters will be verified in the following section.
Indeed, it will be ascertained that there is an infinite sequence of trajectory segments 
in both energy shells with increasing time and hence sequences of functions $t_j(E,E')$, $t'_j(E,E')$
and compound actions $\S_{j,j'}(E, E')$. Thus, combining the complex conjugate contributions of the compound orbits 
with their time reversal, the transition density takes the form
\begin{eqnarray}
P_{EE'}(\tau, \epsilon) \approx & \frac{2^N}{\pi\hbar} \sum_{j,j'} 
\exp\left( -\frac{\epsilon}{\hbar}(|t_j|+|t'_{j'}|)\right)
\\ \nonumber
&\left|\det \frac{\der(E,E')}{\der(t_j,t'_{j'})} ~ \det[\Id - \M(t_j, t'_{j'})] \right|^{-1/2}
\\  \nonumber
&\cos \left[ \frac{1}{\hbar}\S_{j,j'}(E, E')+ \sigma\right] ~.
\label{PSC}
\end{eqnarray}
\begin{figure}
\centering
\includegraphics[width=.6\linewidth]{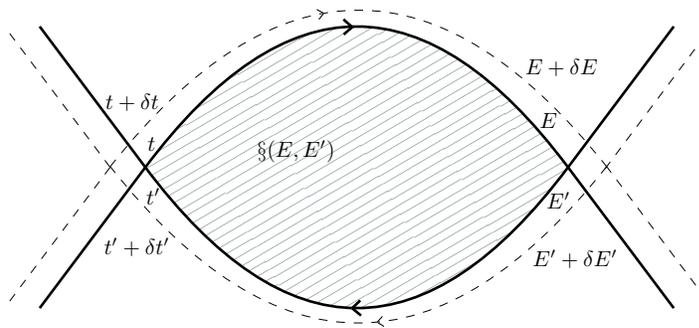}
\caption{The action $\S(E, E')$ of the closed compound orbit is the symplectic area between both orbit segments, which
coincide with the $E$-shell and the $E'$-shell if $N=1$. It is a smooth function of both energies, as are the times $t(E,E')$
and $t'(E,E')$.}
\label{Fig1}
\end{figure}

\section{Sections in energy and traversal times}

The derivatives in the amplitude of the SC contribution of a compound orbit to the energy transition density
imply that it belongs to a continuous family parametrized by $t$ and $t'$. Even though the calculation is carried through 
in the time-domain, this parametrization is most clearly justified in the energy-domain. 
Let us start with the simple case where N=1, so that the compound orbit is identified with segments
of the energy shells $H(\x) = E$ and  $H(\x|\tau) = E'$ joining their intersections, as shown in Fig. 1. 
Evidently, each of these segments in the $E$-shell with $j$ windings 
takes its time $t_j(E,E')$ and likewise the segments on the driven $E'$-shell 
have their duration $t'_{j'}(E,E')$. Then, locally, the inverse function theorem guarantees the existence
of functions $E_j(t,t')$ and $E_{j'}(t,t')$, so that the Jacobian determinant in \eref{PSC} is simply
\be
\det \frac{\der(E_j,E_{j'})}{\der(t,t')} = \left[\det \frac{\der(t_j,t'_{j'})}{\der(E,E')} \right]^{-1}.
\label{Jacinv}
\ee

The extension of this scenario for higher dimensions requires some care. As discussed in {\bf I} and {\bf II},
the pointwise intersection of the pair of energy shells is replaced by a $(2N-2)$-dimensional surface, 
the {\it evolved section}, that is, a generalized Poincar\'e section of the $E$-shell by the driven $E'$-shell, or vice versa. 
Then the $j$th segment determines an orbit in the $j$th Poincar\'e map on the $E$-shell, 
whereas the $j'$th segment corresponds to an orbit on the $j´$'th Poincar\'e map on the driven $E'$-shell.
The compound orbit thus determines a fixed point of the product of the $j$-map with the $j'$-map.

Continuity of this construction for variations of $E$ and $E'$ 
indicate the existence of a two parameter family of $jj'$-compound orbits, 
but this is best nailed down by constructing yet another surface of section. This can be an ordinary Poincar\'e section 
cutting, say the $E$-shell by a fixed $(2N-1)$-D plane. Then a $j$-segment determines an initial point on the fixed section section 
and maps it onto the evolved section. This point is then transported back to evolved section by the continuing $j'$-segment in the $E'$-shell, wherefrom a continuing $j$th segment takes it back to the fixed section. Obviously, a compound orbit determines again a fixed point of this sequence of three mappings, but the advantage is that the energies $E$ and $E'$ are now mere parameters in the product Poincar\'e map. 
If the stability matrix $\mathbf{m}(E,E')$ for the this product map of the $2(N-1)$-D fixed section is nonsingular, 
i.e. $\det[\Id - \mathbf{m}(E,E')] \neq 0$,
then there will be locally a two parameter family of fixed points, corresponding to compound orbits
with the full period $t_j(E,E') + t_{j'}(E,E')$. 

Thus, after all, the situation is just as depicted in Fig. 1, but now, instead of the full $E$-shell and $E'$-shell in the simple case where $N=1$, 
a single $j$-orbit segment in the $E$-shell and a $j'$-orbit segment in the $E'$-shell complete the compound circuit.  
 
For parameters where there is no bifurcation of the closed compound orbits, 
one can use \eref{Jacinv} for all dimensions, so that the contributions of compound orbits 
to the transition density takes the final form 
 \begin{eqnarray}
\fl P_{EE'}(\tau, \epsilon) \approx   \frac{2^N}{\pi\hbar}\sum_{j,j'} \cos \left[ \frac{1}{\hbar} \S_{j,j'}(E, E')+  \sigma\right] ~
\\ \nonumber
\exp\left( -\frac{\epsilon}{\hbar}(|t_j|+|t'_{j'}|)\right)
\left|\det \frac{\der(t_j,t'_{j'})}{\der(E,E')}\right|^{1/2}  ~ |\det[\Id - \M(t_j, t'_{j'})] |^{-1/2}.
\label{PSC2}
\end{eqnarray}
The {jj'}-sum includes all combinations of forward and backward trajectory segments, always obtained as fixed points of the corresponding
generalized Poincar\'e maps, but of course very long orbits will be exponentially dampened by the exponential factors depending on the 
energy width $\epsilon$. One must add to this the classical contributions of 'zero-length' orbits on the intersection of the pair of energy shells,
that is, the evolved section as discussed in {\bf II}, to obtain the complete portrayal of the energy transition density.

\section{Discussion}

The SC approximation of the transition density obtained above can now be compared with that of the previous derivation in {\bf II}.
Resulting from the direct stationary phase approximation of the integral in \eref{PEE1}, over the SC approximation
of the spectral Wigner functions 
\be
\fl W_E(\x,\epsilon) \approx \sqrt{\frac{2}{\pi\hbar}} \sum_j \frac{2^N}{[{\rm d}E/{\rm d}|t_j|~|\det(\Id+ \M(\x,t_j))|]^{1/2}} 
~ \e^{-\epsilon |t_j|/\hbar} ~\cos\left[ \frac{S_j(\x,E)}{\hbar} + \sigma_j \right] ~,
\label{spectral3}
\ee
it can be expressed as
\begin{eqnarray}
\fl P_{EE'}(\tau, \epsilon)  
\approx \frac{2^N}{\pi\hbar}\sum_{j,j'} ~  
\cos \left[\frac{1}{\hbar}\S_{jj'}(E,E') + \sigma_j + \sigma_{j'}+ \frac{\pi}{4}\sigma_{jj'}\right] \\   \nonumber
\exp\left(\frac{\epsilon}{\hbar}(|t_j| + |t_{j'}|) ~ \right) \left|\frac{{\rm d}t_j}{{\rm d}E}~\frac{{\rm d}t_{j'}}{{\rm d}E'}  \right|^{1/2}
  \\   \nonumber
|\det(\Id + \M(\x,t_j) \det(\Id + \M(\x,t_{j'}|\tau))  \det(2\B_j + 2\B_{j'} + \B_j \J \B_{j'} +  \B_{j'} \J \B_j )|^{-1/2}  
\label{SCtrpr}
\end{eqnarray}
(with a correction for the constant factor in {\bf II}).

The action of the closed compound orbit in each term equals that obtained in the present derivation, 
but the same cannot be said for the amplitude. This carried over from the parent spectral Wigner functions,
with the extra factor arising from the stationary phase integral,
which depends on the Hessian matrices, 
\be
\mathbf{B}(\x,t) \equiv \frac{1}{2}~ \frac{\der^2 S(\x,t)}{\der \x^2} ~,
\label{Hessian}
\ee
of the pair of actions combining in the action for the compound orbit.
In all instances, $\x=\x_{jj'}$, is the stationary phase point at the centre of the pair of endpoints
of the pair of trajectory segments composing the {jj'}th compound orbit. 
Thus, the present incorporation of the separate stability matrices within a single determinant, together with the unification of the derivatives
of the times with respect to the energies into a single Jacobian, has somehow incorporated the intricate combination
of Hessian determinants. The spin-off is a considerable simplification of the amplitude.

One may recall that a special choice of coordinates allowed for a factoring into blocks of the stability matrix for a single
orbit segment and hence the absorption of the energy derivative in the amplitude of the spectral Wigner functions in {\bf II}. 
These amplitudes were used in the final amplitude of the transition density contributions in {\bf II}, 
but in retrospect this was not really an advantage. 
The problem is that the special coordinates for factoring the stability matrix depend on the trajectory segment.
So there are two different coordinate systems for the pair, one for each segment. There is no point in
adopting such a procedure for the full compound orbit, because there will be no way to 
include the full Jacobian determinant between the pair of times and the pair of energy derivatives with the full stability matrix.

Semiclassical approximations are generally robust, if one keeps to the lowest orders that are strongly tied to
the underlying classical infrastructure. An example is the equivalence of different quantum representations of the evolution operator
under the stationary phase approximation of the integral transformations between them. In the present instance,
one deals with the probability density of an energy transition, which is an experimental property independent of 
the representation employed in its calculation. The only difference between both approaches is the order in which the trace 
and the Fourier transform is calculated. Considering that the amplitude of the SC oscillations with respect
to changes of parameters $(E,E',\tau)$ is much simpler in its new form, there is no a priory reason not to adopt it.
Their equivalence can then be presumed, even if it seems a hard task to show it directly. In any case,
it is pleasing that the direct derivation in terms of spectral Wigner functions in {\bf II} produces the same phases
as the present double Fourier transform from times to energies.

The present paper closes a sequence which has undergone a somewhat nonlinear development, beyond
adding technical refinement to semiclassical formulae for general transitions between coarsegrained energy levels.  
The initial paper {\bf I} expanded an identity between pure Wigner functions, while bestowing it physical significance,
though its main results did not go beyond purely classical energy transitions in a narrow context. This was followed in {\bf II} by the
addition of quantum terms to this classical background, which oscillate with the variation of the transition parameters,
either the transition energies, or, for instance, the the duration of the external driving in an expanded context that included Hamiltonian evolution. The phase of the oscillations were identified with the action of closed compound orbits, composed of orbit segments in the in the initial and final energy shells, but their amplitude, derived by the stationary phase approximation of the original Wigner identity, was far from transparent.

It is only here, through the present rederivation of the transition density, that the amplitude of the semiclassical contribution of the compound orbits follows their phase in achieving full independence of the chosen Wigner-Weyl representation. In this way, 
the contributions of closed compound orbits for dynamical energy transitions, based on their classical actions and stability matrices, emulates
the contributions of periodic orbits in a single energy shell to the static energy spectrum, in the manner disclosed by Gutzwiller's ground breaking work on the trace formula \cite{Gutzbook}. Again, this work also deals with a trace, but it involves a product of operators, which leads to the replacement of simple periodic orbits by closed compound orbits.

The difficulty of working directly with these semiclassical formulae based on many trajectories delayed significant numerical verification of formulae for the spectrum for a long time, but the fact that here one is addressing coarsegrained energy shells and, hence,
cutoffs in the duration of the closed compound orbits, allows for the possibility that numerical simulations may be forthcoming.

\section*{Acknowledgments}
I thank Gabriel Lando for his help in preparing the figure.
Partial financial support from the 
National Institute for Science and Technology--Quantum Information
and CNPq (Brazilian agencies) is gratefully acknowledged.

\end{document}